\documentclass{ws-ijmpb}
\usepackage{float,color}
\definecolor{Blue}{rgb}{0,0,0.63}
\definecolor{Red}{rgb}{1,0,0}
\definecolor{Green}{rgb}{0,0.5,0}
\definecolor{Purple}{rgb}{0.5,0,1}
\definecolor{Yellow}{rgb}{1,1,0.5}
\definecolor{Cyan}{rgb}{0.5,1,1}
\definecolor{Grey}{rgb}{0.25,0.25,0.25}

\begin{document}

\title{Extraordinary Magnetoresistance in Hybrid\\Semiconductor-Metal Systems}

\author{T.H. Hewett and F.V. Kusmartsev}
\address{Department of Physics, Loughborough University, Loughborough, LE11 3TU, UK}

\maketitle


\begin{abstract}
We show that extraordinary magnetoresistance (EMR) arises in systems consisting of two components; a semiconducting ring with a metallic inclusion embedded. The important aspect of this discovery is that the system must have a quasi-two-dimensional character. Using the same materials and geometries for the samples as in experiments by Solin {\it et al.}\cite{Solin1,Solin2}, we show that such systems indeed exhibit a huge magnetoresistance. The magnetoresistance arises due to the switching of electrical current paths passing through the metallic inclusion. Diagrams illustrating the flow of the current density within the samples are utilised in discussion of the mechanism responsible for the magnetoresistance effect. Extensions are then suggested which may be applicable to the silver chalcogenides. Our theory offers an excellent description and explanation of experiments where a huge magnetoresistance has been discovered\cite{Solin2,Xu2}. 
\end{abstract}

\keywords{Magnetoresistance; Current paths; Strongly inhomogeneous.}

\section{Introduction}

The magnetoresistance effect known as extraordinary magnetoresistance (EMR) was coined by Solin {\it et al.}\cite{Solin2} after the experimental discovery in 2000. This effect showed extremely large room temperature magnetoresistance values of a million percent in a 5T applied magnetic field\cite{Solin1}. These values were found in composite van der Pauw disks consisting of a conducting and semiconducting region of non-magnetic materials. As a result of the observed properties of EMR it has been proposed that improvements could be achieved in the read heads of magnetic disk drives\cite{Solin3,Solin5}.

In 1997 Xu {\it et al.}\cite{Xu2} discovered a large magnetoresistance in non-magnetic silver chalcogenides ${Ag}_{2+\delta}Se$ and ${Ag}_{2+\delta}Te$. The addition of a small excess amount ($\delta$) of silver atoms into the structure of the semiconductors  ${Ag}_{2}Se$ and ${Ag}_{2}Te$ (with no appreciable magnetoresistance themselves) caused a large linear magnetoresistance to be observed. At room temperature and in a magnetic field of 5.5T the magnetoresistance was 200\% with no sign of saturation. The magnetoresistance displayed a linear dependance on magnetic field which remained in low fields\cite{Xu2}. A model for this magnetoresistance has been proposed\cite{Milton,Bulgad4,Xu} which considers the silver chalcogenides as consisting of two components: a semiconductor and a conducting component. This is analogous to the EMR effect in systems where it may arise. Bulgadaev and Kusmartsev\cite{Bulgad3,Bulgad2} have found explicit expressions for the magnetoresistance of strongly inhomogeneous two-phase systems by developing a method of conformal mapping transformations. This method utilises the exact dual transformation in relating the effective conductivity of plannar inhomogeneous two-phase systems with and without an applied magnetic field. They have studied three models: the random droplet model (RDM); the random parquet model (RPM); and an effective medium model (EMM). 

Parish and Littlewood\cite{Parish2} have modelled a strongly inhomogeneous conductor using a random resistor network made from four terminal resistors, and have shown that the properties of the magnetoresistance are similar to those of the silver chalcogenides. Abrikosov\cite{AAA3} proposed an explanation of the abnormal magnetoresistance observed in the silver chalcogenides, namely quantum magnetoresistance, based on the assumption that in such systems gapless dirac fermions arise. Recent experimental work has verified that the magnetoresistance effect in the silver chalcogenides is somehow related to the microstructure of the material, which has been identified and investigated\cite{Janek1,Janek3}.

In this paper van der Pauw disks consisting of a conducting and semiconducting region of non-magnetic materials are considered. Their electrical properties, such as conductivity and the distribution of the electric field and current, are studied using finite element simulations. These were based on the EMR geometry used by Solin {\it et al.}\cite{Solin2} when the effect was first discovered. The experimental results are verified and the mechanism for the effect discussed with the use of streamline plots of total current density. The paper is structured as follows: firstly, a description of the approach taken in order to produce the simulations including deviations from the experimental geometry; secondly, the magnetoresistance results are given along with the streamline plots; and finally, discussion of the results in comparison with previous experiments, with the mechanism for the magnetoresistance offered as well as proposals for extension.

\section{Simulation approach}
\label{sec:Simulation approach}

We modelled the experimental construction of the composite van der Pauw disk, as used by Solin {\it et al.}\cite{Solin2}, in the discovery of the EMR effect. The experimental construction consisted of a circular disk of Indium Antimonide (InSb) of radius $r_b = 0.5mm$ with a concentric inclusion of Gold (Au) of radius $r_a$. Different geometries were achieved by varying the radius of the Au inclusion and quantified with the use of the filling factor $\alpha$. 

\begin{equation} \label{eq:alpha}
\alpha=\frac{r_a}{r_b}=\frac{n}{16}
\end{equation}

Alternatively the volume fraction (f), the fraction of conducting to semiconducting material, can be defined as follows.

\begin{equation} \label{eq:volfrac}
f=\frac{\pi{r_a}^2}{\pi{r_b}^2}={\alpha}^2
\end{equation}

Experimentally disks were grown on a Gallium Arsenide (GaAs) substrate, with a $1.3\mu m$ thick active layer of InSb. It was noted that the side walls between the InSb and the Au regions were approximately ${19}^{\circ}$ from vertical\cite{Solin2}. The magnetoresistance was determined utilising the van der Pauw method of measuring the resistivity of a thin film\cite{Pauw2}. This required four Au contacts to be placed equidistant around the perimeter of the disk. Two adjacent contacts were used to inject the current into the system, with the other two contacts used to measure the resulting potential difference. To measure the magnetoresistance the magnetic field was applied perpendicular to the surface of the disk.

\begin{figure}[t]
	\centering
	\includegraphics[width=0.6\textwidth]{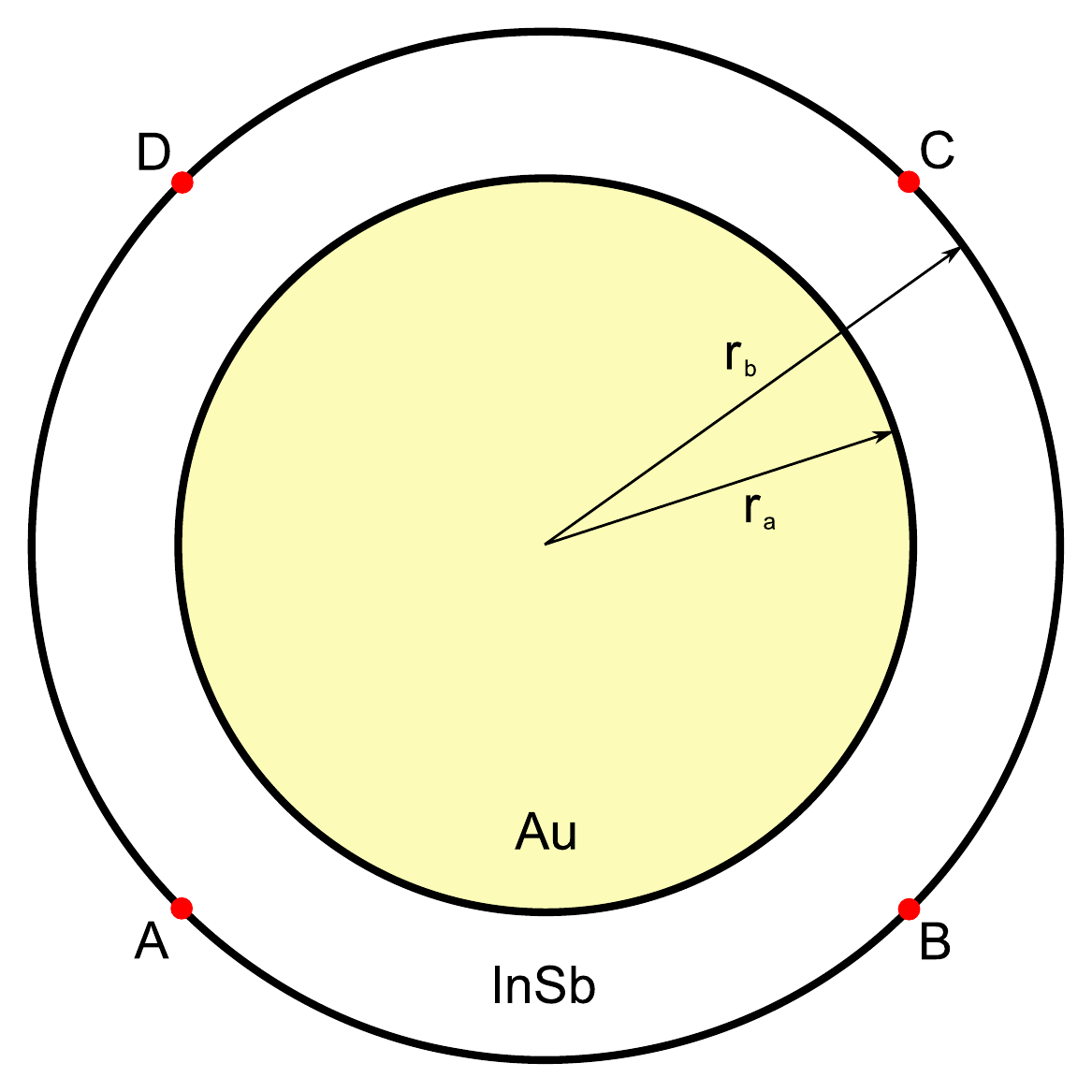}
	\caption{The geometry of the simulated system, illustrating the semiconducting ring (of radius $r_b$) with a conducting inclusion (of radius $r_a$), including the four point contacts. Here the current is flowing through contacts A and B, while the voltage is measured across contacts C and D.}
	\label{fig:Geom2}
\end{figure}

In order to simulate this system some aspects were required to be adapted. The first simplification made was to simulate the system in two dimensions. This was justified since the experimental disk was of a thin film construction. Consequently, experimental errors in the angle of the side walls between the two regions were neglected. Additionally, only the active layer of the disk was considered with all other layers of experimental significance ignored for the purposes of simulation. The four contacts on the disk were assumed to be point contacts with experimental errors disregarded. The simulations were carried out using the finite element analysis software package COMSOL multiphysics. With such assumptions the two-dimensional geometry used for simulations is given in Fig. \ref{fig:Geom2}.

Various geometries have been considered, with the filling factor varying by $1/16$ of $r_b$ from $0$ to $15/16$. With the geometry considered, various parameters were defined and quantified. The simulations were carried out with an applied field ranging from 0 -- 5T mirroring those used experimentally\cite{Solin2}. In order to measure the magnetoresistance an electrical current was applied through contacts A and B. Using the van der Pauw method for measurement of the resistivity leads to the following expression of the magnetoresistance.

\begin{equation} \label{eq:MR}
MR=\frac{R(H)-R(0)}{R(0)}=\frac{{V}_{CD}(H)}{{V}_{CD}(0)}-1
\end{equation}

Where ${V}_{CD}(H)={V}_{C}(H)-{V}_{D}(H)$. In order to specify the required materials the conductivities and charge carrier mobilities were utilised to establish the conductivity tensor for each of the two regions. The conductivity and mobility values used were as close to the experimental system as possible. The values of the conductivity used for InSb and Au were $\sigma_{InSb}=1.86\times10^{4} (1/{\Omega m})$ and $\sigma_{Au}=4.52\times10^{7} (1/{\Omega m})$ respectively. With the corresponding mobilities of $\mu_{InSb}=4.55 ({m}^{2}/Vs)$ and $\mu_{Au}=5\times10^{-3} ({m}^{2}/Vs)$. Using these values, and the parameter $\beta=\mu H$, the conductivity tensor could be defined for each of the two materials using the following expression.

\begin{equation} \label{eq:tensor}
\hat{\sigma}=
\left( \begin{array}{cc}
\sigma_{xx} & \sigma_{xy} \\
\sigma_{yx} & \sigma_{yy} 
\end{array} \right)
=
\left( \begin{array}{cc}
\frac{\sigma}{1+\beta^2} & \frac{-\sigma \beta}{1+\beta^2} \\
\frac{\sigma \beta}{1+\beta^2} & \frac{\sigma}{1+\beta^2} 
\end{array} \right)
\end{equation}

Finally the boundary conditions were specified. On the perimeter of the entire disk (the semiconducting region) they were given as electrical insulation, while at the interface between the semiconducting ring and the conducting inhomogeneity they were set to continuity. These simulations were carried out with a mesh consisting of approximately 60,000 triangular elements, with the mesh being more refined where the variation in the potential was greatest.

\section{Results}
\label{sec:Results}

\begin{figure}[t]
	\centering
	\includegraphics[width=0.8\textwidth]{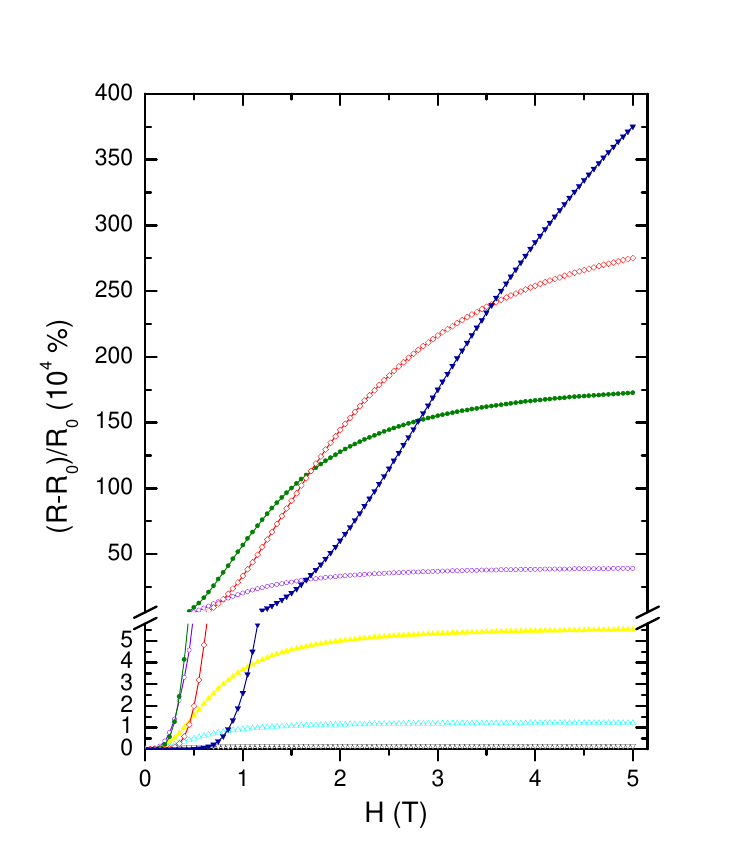}
	\caption{The magnetoresistance as a function of applied magnetic field. Here the symbols correspond to various geometries: n = 15 ({\color{Blue}{$\blacktriangledown$}}), 14 ({\color{Red}{$\diamond$}}), 13 ({\color{Green}{$\bullet$}}), 12 ({\color{Purple}{$\circ$}}), 11 ({\color{Yellow}{$\blacktriangle$}}), 10 ({\color{Cyan}{$\vartriangle$}}) and 8 ({\color{Grey}{$\triangledown$}}). The largest value of magnetoresistance seen at a field of 5T corresponds to a geometry with n = 15, with the lowest value of magnetoresistance at 5T corresponding to a geometry with n = 8.}
	\label{fig:MRvH}
\end{figure}

Fig. \ref{fig:MRvH} presents the magnetoresistance as a function of applied magnetic field. Here we see the magnitude of the magnetoresistance reaching 3,750,000\% in a geometry with $n=15$ ($f=0.879$) and field of 5T.

The magnetoresistance calculated has an extremely large magnitude already at a 5T magnetic field and can be realised at room temperatures. Systems with small conducting inclusions ($n = 1$ -- $8$) show saturation in the magnetoresistance at a magnetic field of approximately 1.5T. Larger inclusions require a larger magnetic field to reach their maximum magnetoresistance value and thus saturation occurs at higher magnetic fields. The geometries with $n = 14$ and $15$ do not exhibit saturation of the magnetoresistance in a field of up to 5T. This indicates that the saturation field increases with the size of the conducting region.

The form and magnitude of the simulated results are in agreement with experimental data\cite{Solin2}, with the values especially close at lower applied fields. The magnetoresistance being found to be $79\%$ with $n=12$ ($f=0.563$) in a field of 0.05T and $5721\%$ with $n=13$ ($f=0.660$) in a 0.25T applied field. 

In a quasi-two-dimensional system it is important that the conducting inclusion extends through the entire thickness of the structure. With the application of a magnetic field, perpendicular to the surface of the system (z direction), only charge carrier motion in the x-y plane is affected. Charge carriers travel with a motion unaltered by the magnetic field when their trajectory is in a parallel direction (along the z axis). Therefore, if the conducting inclusion does not extend the entire thickness of the structure alternate current paths are created (the current can avoid the conducting inclusion by travelling underneath). This causes a dramatic reduction in the magnitude of the EMR effect. 

The magnitude of the simulated magnetoresistance is found to be larger than that found experimentally for $n=14$ ($f=0.766$) and $15$ ($f=0.879$) especially in high fields of approximately 5T. This difference may be a result of the experimental system containing side walls between the semiconducting and conducting region that deviate from vertical. However, very good agreement is found for smaller Au inclusions, for example a magnetoresistance of $39\times10^4\%$ for $n=12$ ($f=0.563$) and $173\times10^4\%$ for $n=13$ ($f=0.660$) both in a field of 5T.

\begin{figure}[t]
	\centering
	\includegraphics[width=0.8\textwidth]{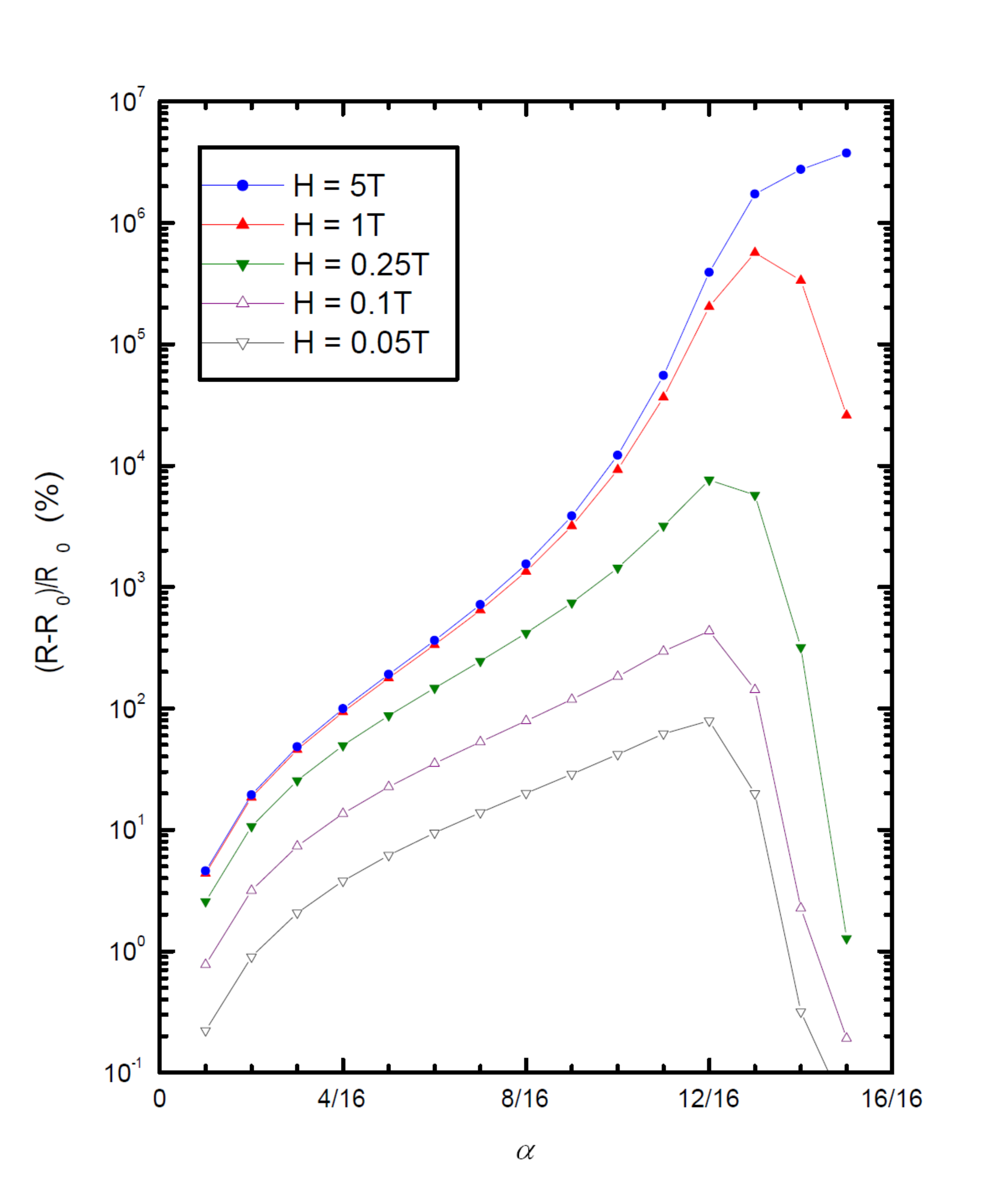}
	\caption{The magnetoresistance as a function of filling factor ($\alpha$) at 5 various applied magnetic fields: H = 0.05T ({\color{Grey}{$\triangledown$}}), 0.1T ({\color{Purple}{$\vartriangle$}}), 0.25T ({\color{Green}{$\blacktriangledown$}}), 1T ({\color{Red}{$\blacktriangle$}}) and 5T ({\color{Blue}{$\bullet$}}) }
	\label{fig:MRvAlpha}
\end{figure}

Fig. \ref{fig:MRvAlpha} shows the simulated magnetoresistance as a function of filling factor, $\alpha$. This plot shows a good agreement to the experimental results\cite{Solin2} yet some differences are apparent. We notice that in general the magnetoresistance increases with $\alpha$ up to a certain value, with the value of $\alpha$ at which the magnetoresistance peaks increasing with higher applied fields. At low magnetic fields (H = 0.05, 0.1 and 0.25T) the magnetoresistance peaks at $n=12$ and for higher fields (H = 1 and 5T) peaking at $n= 13$ and $15$. One may notice that the data for the field of 5T does not fall after the peak magnetoresistance occurs. Instead the magnetoresistance is its largest for $n=15$, the geometry with the largest Au inclusion. Experimentally it was found that the magnetoresistance dropped after $n=13$ in a 5T field and did not continue to increase. This difference may be a result of experimental uncertainties, especially those errors in the contacts deviating from ideal point contacts. The errors associated with the contacts are more pronounced in geometries where the semiconducting ring is narrow (large values of $\alpha$).

\begin{figure*}[t]
	\centering
	\includegraphics[width=\textwidth]{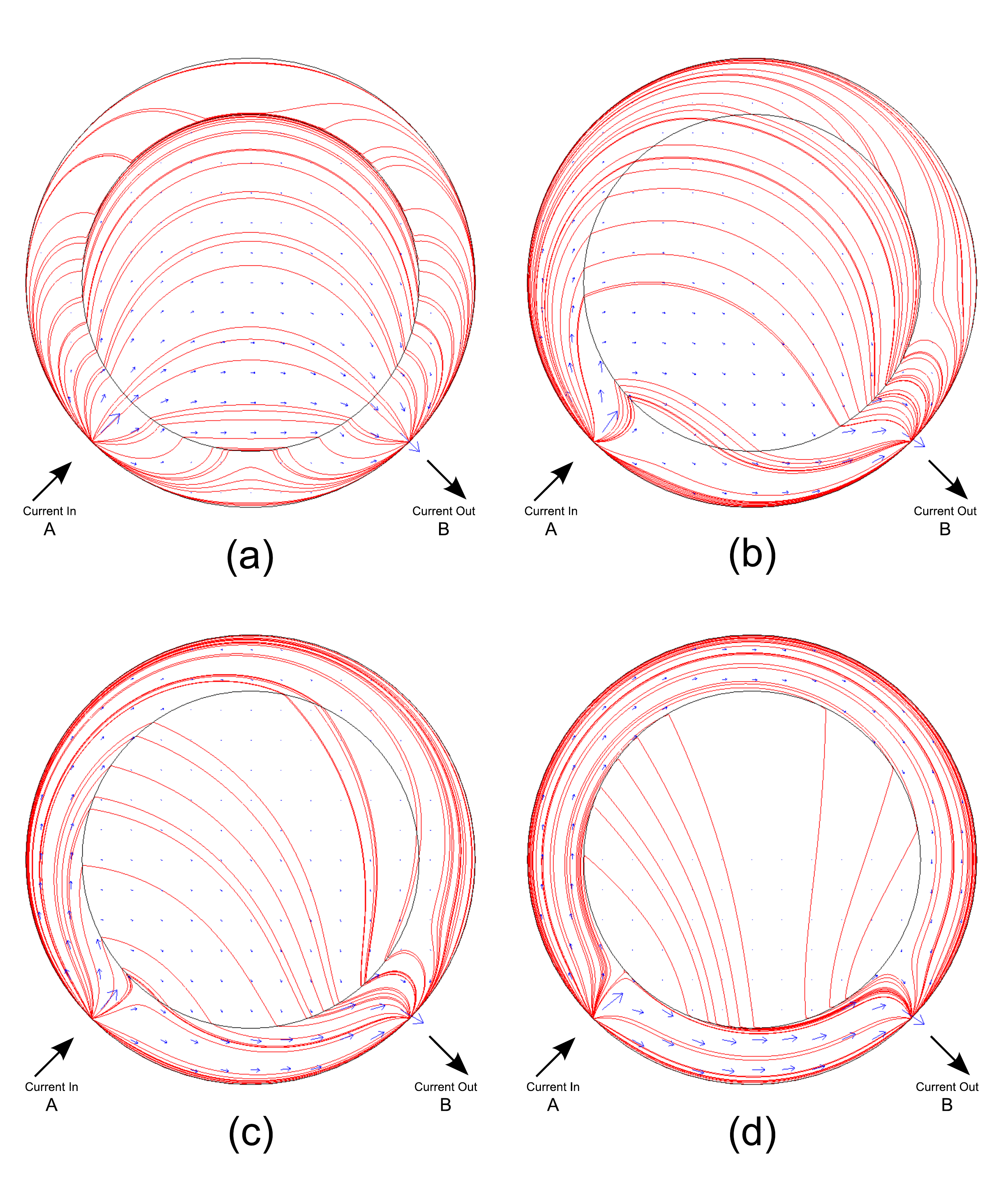}
	\caption{Visualisations of the path of the total current density in a disk where $n=12$ ($f=0.563$). The solid streamlines (red) indicate lines of the same current density, while the arrows (blue) represent the relative magnitude and direction of the current flow at the given points in the system. Figures (a) -- (d) show the effect of an applied magnetic field on the flow of current. (a) $H=0$, (b) $H=0.5$, (c) $H=1$ and (d) $H=5$.}
	\label{fig:n12Iflow}
\end{figure*}

We represent the path of the total current density for a single system where $n=12$ (the concept generally holds for other geometries) at various applied magnetic fields in Figs. \ref{fig:n12Iflow}(a) -- (d). The streamlines (red) link points of the same current density while the arrows (blue) show the magnitude and direction of the current flow at that particular point in the system. These diagrams shows the current injected at point A and taken off at point B.

In Fig. \ref{fig:n12Iflow}(a) we see the path of the current flow throughout the system with zero applied magnetic field. The current flow here is predominantly directed into the Au inclusion, which can be seen most clearly from the arrows (blue), with very little of the total current flowing around the perimeter of the disk in the semiconducting region. At zero magnetic field the electric field lines are perpendicular to the equipotential Au surface. With no applied magnetic field the current flow is parallel to the electric field lines and so flows in the same direction, into the Au inclusion. This flow into the conducting region constitutes a system of low resistance.

In the next diagram Fig. \ref{fig:n12Iflow}(b), the applied magnetic field is 0.5T. The current is still predominantly flowing through the Au inclusion however the magnetic field has had the effect of distorting the current path. Now a larger proportion of the current is being forced to flow around the semiconducting region of high resistance. The magnetic field causes the electric field (which is still perpendicular to the surface of the Au inclusion) and the current flow to stray away from being parallel with the electrical field lines. The angle between the electric field and the current (Hall angle) increases with higher applied magnetic fields.

There is a continuation of this trend in Fig. \ref{fig:n12Iflow}(c) where a larger magnetic field ($H=1T$) has caused the angle between the electric field and the current flow to increase to near $90^\circ$. This means the current flow and electric field are almost perpendicular. Thus the current is directed away from the conducting region and forced to flow through the semiconducting outer ring of much higher resistance. We see here that most of the current (arrow plot) is travelling from point A to point B through the outer ring of the system and avoiding the lower resistance Au path.

Finally, with a high applied magnetic field of $H=5T$ represented in Fig. \ref{fig:n12Iflow}(d) the directions of the current flow and the electric field are almost perpendicular. Now the majority of the current is flowing around the Au inhomogeneity in the region of semiconductor material. The application of the magnetic field causes the current to avoid the low resistance path and forces it to flow along narrow (depending on the value of $\alpha$) channels of semiconductor. This results in a system with a substantially higher resistance than when no magnetic field is applied. This effect causes the huge geometric magnetoresistance known as EMR that was found in experiments\cite{Solin2}.

The geometry of the system plays an important role in the magnetoresistance effect. With a relatively small Au inclusion a large proportion of the current already flows through the wide semiconducting channel at $H=0$. With the application of the magnetic field in such geometries a small proportion of the current switches from flowing through the Au to through the semiconductor. This leads to smaller values of magnetoresistance with low values of $\alpha$. With large Au inclusions however, more of the current initially flows through the Au and thus gets switched to flow through the semiconductor when a magnetic field is applied. More of the current is affected by the application of the magnetic field, therefore larger values of magnetoresistance are observed. Additionally in geometries with large conducting inclusions the semiconducting channels are narrow, this increases the resistance of current flow through the semiconductor material, thus further enhancing the magnetoresistance. 

In geometries with large conducting droplets a larger field is required to force the current to flow through the narrow semiconducting regions. For this reason the geometries with larger values of $\alpha$ peak in their values of magnetoresistance at higher applied fields. This explains the trend of the peak magnetoresistance value occurring at higher values of $\alpha$ at higher magnetic fields observed both here (Fig. \ref{fig:MRvAlpha}) and in experimental data\cite{Solin2}.

These current flow diagrams reinforce ideas discussed by Solin {\it et al.}\cite{Solin1} concerning the EMR effect. They regard the system as a short circuit (current flowing through the conducting region) when the magnetic field is small and an open circuit (current deflected around conducting region) when the magnetic field is large.

\begin{figure}[t]
	\centering
	\includegraphics[width=0.92\textwidth]{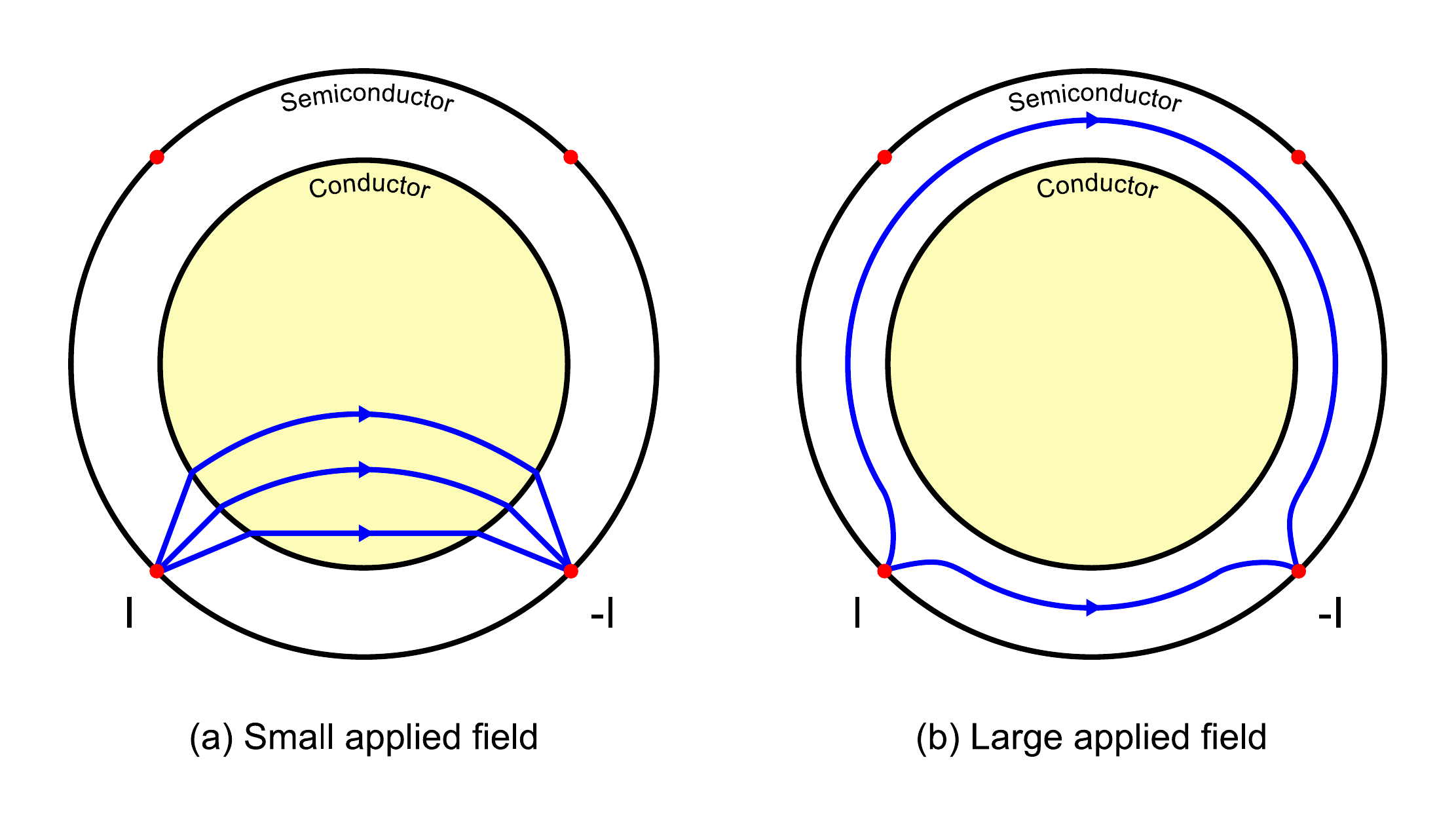}
	\caption{Simplified illustration of the current flow in a typical system in both low and high applied magnetic fields}
	\label{fig:Iflow}
\end{figure}

The application of the magnetic field can be thought of as a switch; with zero or a small applied field the majority of the current flows through the conducting Au inclusion giving a low resistance path. However, the application of a large magnetic field forces the current to flow around the perimeter of the disk avoiding the conducting inclusion, as if there was a ring of semiconductor with a cavity in the centre, thus creating a significantly higher resistance path, see Fig. \ref{fig:Iflow}.

\section{Further Discussion}
\label{sec:Further Discussion}

The EMR effect strongly depends on the geometry of the system. The physical component to the magnetoresistance (ordinary magnetoresistance) being almost negligible for both the conducting and semiconducting materials individually. Other large magnetoresistance effects originate from a large physical component that relies on the electrical properties of the materials involved. The EMR effect may therefore be applicable to many different systems, where droplets or regions of material with a significantly higher conductivity to that of the surrounding material are observed.

Due to the high economic value of Au, practical devices based on EMR would be more cost effective to produce if the EMR effect was efficient and produced the highest value of magnetoresistance with a fixed quantity of Au. In order to optimise the amount of Au required the geometry could be modified, for example, with the implementation of different shaped droplets (oval shaped droplets instead of circular ones)\cite{phkr}.

Here the effect of a single conducting droplet embedded in a semiconductor is investigated, by replicating the experimental geometry in which the EMR effect was discovered. However, further research will be carried out investigating the effect of multiple conducting droplets embedded in a semiconducting material. Simulation of a random droplet model may mean the magnetoresistance follows a linear growth with magnetic field as is desirable practically. A multiple droplet system may be hard to create precisely but some systems form droplets of conducting and semiconducting material spontaneously. An example of this is the silver chalcogenides which have shown a large room temperature magnetoresistance, where a small excess of conductor is added to a semiconducting material. The addition of this small excess amount of conducting material has been shown to form a percolating silver network along the grain boundaries (for already low silver excess) and even conducting droplets in the microstructure of the material (for higher excess silver content)\cite{Janek3} in analogy with this work. It would be interesting to look at a random parquet model as a parquet type structure has also been identified\cite{Janek1}.

\section{Conclusion}
\label{sec:Conclusion}

Since this paper has been produced a similar piece of work has come to our attention, namely that of Moussa {\it et al.}\cite{Moussa}, which documents finite element simulations regarding the EMR effect. Their results show the field dependence of the current flow, the potential on the disk periphery and compare the disk resistance with experiment. These simulations were carried out with a mesh consisting of 6000 nodal points. The results given are in excellent agreement with those presented in this paper and with the experimental data\cite{Solin2}. Our results were calculated using a much more sophisticated mesh with the magnetoresistance explicitly calculated and compared to the experimental data.

We have investigated the EMR effect using finite element simulations. The simulated geometry was considered in two-dimensions with parameters described in experiments by Solin {\it et al.} so as to replicate these systems. The magnetoresistance was calculated for different geometries ($n=1$ -- $15$) and in a range of magnetic fields (H = 0 -- 5T).

We have found that a huge magnetoresistance appears, due to the geometry of the system, and our results show good agreement with the experimental values. The size of the conducting inclusion is found to affect the magnetoresistance greatly. We have looked at diagrams of the total current density flow throughout the samples at various applied magnetic fields and discussed the mechanism for the resulting huge magnetoresistance. This model can now be utilised in the investigation of the effect of various droplet shapes on the magnetoresistance.


\section*{References}
\bibliographystyle{unsrt}
\bibliography{MRref}

\end{document}